\documentstyle[12pt]{article}

\def\tilde{\widetilde}
\def\hat{\widehat}

\newcommand{\prd}{Phys.\ Rev.\ D }

\newcommand{\pl}{Phys.\ Lett.\ }

\newcommand{\np}{Nucl.\ Phys.\ B }
\newcommand{\sij}{\sigma_{ij}}
\newcommand{\ha}{\frac{1}{2}}
\newcommand{\pam}{\partial_\mu}
\newcommand{\Amnl}{A_\mu^{nl}}
\newcommand{\snl}{\sigma_{nl}}
\newcommand{\bps}{\bar\psi_r}
\newcommand{\ps}{\psi_r}
\newcommand{\bph}{\bar\phi_r}
\newcommand{\ph}{\phi_r}
\newcommand{\btph}{\overline{\tilde\phi}_r}
\newcommand{\tph}{\tilde\phi_r}
\newcommand{\pimn}{\Pi_{\mu\nu}}
\newcommand{\g}{\gamma_5}
\newcommand{\gm}{\gamma_\mu}
\newcommand{\gn}{\gamma_\nu}
\begin{document}
\hfill MPI-Ph 93-12\newline\mbox{ }
\hfill March 1993

\begin{center}
{\large\bf Removing fermion doublers in chiral gauge theories\\
on the lattice}\\
\vspace{1cm}
{\sc S.A. Frolov}\footnote{LPTHE. Tour 16, 1-er etage,
Universite Pierre et Marie Curie 4; Place Jussieu 75252
Paris CEDEX 5, France, on leave from Steclov Mathematical
Institute, Moscow. e-mail: frolov@lpthe.jussieu.fr.}

and

{\sc A.A. Slavnov}\footnote{On leave from Steclov Mathematical
Institute, Vavilov 42, GSP-1, 117966, Moscow, Russia,
e-mail"slavnov@qft.mian.su}\\

\vspace{1cm}
{\sl Max-Planck-Institut f\"ur Physik,\\
Werner-Heisenberg-Institut f\"ur Physik\\
F\"ohringer Ring 6, 8000 M\"unchen 40, Germany\\
}
\end{center}
\bigskip
\begin{abstract}
A method for removing fermion doublers in anomaly free chiral
gauge theories on the lattice is proposed. It is proven that the
resulting continuum theory is gauge invariant and does not require
noninvariant counterterms of fine tuning of parameters.
\end{abstract}
\clearpage
\setcounter{page}{1}
\pagestyle{plain}
\section{Introduction}
In this paper we propose a method for removing fermion doublers
in anomaly free chiral gauge theories which preserves the gauge
invariance in the continuum limit. This problem is closely
related to the problem of constructing an explicitely invariant
regularization for anomaly free chiral gauge models in the
continuum case. In spite of the numerous efforts
\cite{dre}-\cite{ban}, this problem for a long time
had no satisfactory solution although some proposals
still need further investigations \cite{bor}-\cite{zen}.
(See also the reviews \cite{smi}, \cite{gol} for more
complete references.) Moreover a ``no go'' theorem
has been proven \cite{nie} stating that under some
plausible conditions such a regularization cannot
exist. However, recently it was shown that this ``no go'' theorem
may be avoided. In our paper \cite{fro} the explicitely
invariant regularization for the continuum $SO(10)$ model
as well as for the standard model was constructed with the
help of the infinite series of the Pauli-Villars (PV)
fields. At the same time D.~Kaplan \cite{kap} proposed
the lattice formulation of chiral gauge models based on
the introduction of the extra dimension. In this model
all the doublers may be given large invariant masses. It
was argued that for anomaly free models this construction
leads to an acceptable gauge invariant continuum theory
although the proof still has to be given.

It was indicated in the paper \cite{nar} that the Kaplan's
procedure is in fact also equivalent to introducing infinitely
many regulator fields and in this sense both proposals
\cite{fro,kap} use a similar mechanism.

In this paper we shall show that the method
proposed in our paper \cite{fro} can be generalized to
lattice gauge models. The regularization which will be
described below breaks a manifest gauge invariance for a finite
lattice spacing. However, it will be proven that in the
continuum limit the gauge invariance is restored and no
noninvariant counterterms or fine tuning are needed.

The idea to use a regularization breaking the gauge
invariance for a finite lattice spacing was discussed
before \cite{bor,alo}, but in general such a procedure
requires gauge noninvariant counterterms and fine tuning
of the parameters. In our approach this problem is absent.

The paper is organized as follows. In section~2 we discuss the
$SO(10)$ model with an even number of generations. This case
is technically simpler and allows to present the main
idea in a more transparent way. We firstly remind the construction
of the invariant regularization for the continuum model
and then describe the lattice formulation which leads in the
continuum limit to the gauge invariant theory without doublers.
Section~3 is devoted to the discussion of the $SO(10)$ model
and the standard model with the odd number of generations.
\vskip0.8cm
\section{SO(10) model with an even number of generations}

We start by reminding the main idea of the regularization
proposed in our paper \cite{fro}.

The unified $SO(10)$ model may be described by the
Lagrangian
\begin{equation}
{\cal L} =-\frac{1}{4}(F_{\mu\nu}^{ij})^2+i\sum_{k}\overline{\psi}^k_+
(\partial_\mu -ig A_\mu^{ij}\sigma_{ij})\psi^k_+
\label{F1}
\end{equation}
Euclidean space formulation is used although it is not essential.
We use the notations of ref.~ \cite{fwil}. Here $\psi_+$
are the 16-component chiral spinors describing quark and
lepton fields. Index $k$ numerates generations. The
matrices $\sigma_{ij}$ are defined as follows:
$\sij = \frac{1}{2}\lbrack\Gamma_i,\Gamma_j\rbrack$,
where $\Gamma_i$ are Hermitian 32$\times$32
matrices which satisfy the Clifford algebra:

\begin{equation}
\lbrack\Gamma_i,\Gamma_j\rbrack_+ =\delta_{ij}
\label{F2}
\end{equation}
The following equation holds
\begin{eqnarray}
U^{-1}(R)\Gamma_k U(R)&=&R_{\kappa l}(\omega)\Gamma_e\nonumber\\
U(R)&=& exp\{i\omega_{kl}\sigma_{kl}\}\/,\label{F3}
\end{eqnarray}

where the matrices $R_{kl}(\omega)$ determine a rotation in
a 32-dimensional space.

The mapping $R\to U(R)$ defines a 32-dimensional representation
of SO(10). This representation is reducible and to construct the
irreducible 16-dimensional representation one uses the ``chiral''
projections

\begin{equation}
\psi_\pm =\ha (1\pm \Gamma_{11})\psi\,,
\label{F4}
\end{equation}
where
\begin{equation}
\Gamma_{11}=-\Gamma_1\Gamma_2\cdots \Gamma_{10}
\label{F5}
\end{equation}

We assume also that the spinors $\psi_+$ are Weyl spinors:
$\psi_+\equiv\ha (1+\gamma_5)\psi_+$.

In the following we shall consider only the regularization
of spinorial loops having in mind that the Yang-Mills fields may
be regularized in a gauge invariant way by using the higher covariant
derivative method \cite{sla}. For definiteness the number of
generations is chosen to be equal to 2.

We shall use for the regularization of the spinorial loops some
modification of the Pauli-Villars method introducing the
interaction with the auxiliary fermionic spinor fields $\psi_r$
and the bosonic spinors $\phi_r$. The usual
obstacle for using the Pauli-Villars method in chiral theories
is the impossibility to introduce the mass term for the
auxiliary fields as the combination $\bar{\psi}\psi$ is
equal to zero for Weyl spinors. However in our case the
Majorana mass term may be introduced due to the
existence of a ``conjugation'' matrix $C$, satisfying the
relation

\begin{equation}
\sij^T C=-C\sij\,.
\label{F6}
\end{equation}
The expression
\begin{equation}
M_r(\psi_r^T C_D C\Gamma_{11}\psi_r +\,\mbox{h.c.})\,,
\label{F7}
\end{equation}

where $C_D$ is the usual charge conjugation matrix, provides
a gauge invariant mass term for the $P-V$ fields. However,
in this eq.\ $\psi_r$ are the 32-component spinors
$\psi_r=\psi^r_++\psi^r_-$ and not the 16-component ``chiral''
ones. One cannot write a gauge invariant mass term using
only positive chirality spinors as the combination
$\psi^T_+ C C_D\Gamma_{11}\psi_+$ is identically zero.

At first sight we meet again the same problem: the original
theory includes only positive chirality spinors and to introduce
the $P-V$ fields we need both positive and negative chirality
spinors. The crucial observation is that for the SO(10) model the
positive and negative chirality SO(10) spinors give the same contribution
to the divergent diagrams. The difference between the positive and
negative chirality sponors arises only in diagrams with the
number of external lines $>4$, which are convergent.

Al\-ter\-na\-tive\-ly,\, re\-pla\-cing
the right-handed spi\-nors by charge con\-ju\-ga\-ted\, left-handed
ones one can see that the regularization
is needed only for parity conserving part of the diagrams.
The parity odd part is different from zero only for the
convergent diagrams. This property holds also for the
standard model and presumably for all anomaly free chiral
models. Let us prove it for the model under consideration.

The interaction vertex in (\ref{F1}) includes the projection
operator $\ha (1+\Gamma_{11})$. Calculating a spinorial
loop one should take a trace

\begin{equation}
Tr (\sij\sigma_{kl}\ldots (1+\Gamma_{11}))
\label{F8}
\end{equation}

Due to the definition of $\Gamma_{11}$, (\ref{F5}), the
trace

\begin{equation}
Tr(\sigma_{i_1j_1}\sigma_{i_2j_2}\ldots\sigma_{i_nj_n}\Gamma_{11})
\label{F9}
\end{equation}

is equal to zero if $n<5$. Therefore for the divergent diagrams
with $n\le 4$ the
part proportional to $\Gamma_{11}$ vanishes and the
positive and negative chirality spinors give the same contribution.
The total contribution of the 32-component spinor
$\psi =\psi_+ +\psi_-$ to the divergent diagrams
is twice as big as the contribution of the 16-component
``chiral'' spinor. Hence if the number of generations
in the original Lagrangian (1) is equal to 2, one can
use for its regularization the 32-component spinors
$\psi_r$ with the mass term (\ref{F7}).

Having this in mind one can take as a regularized Lagrangian
the following expression:
\begin{eqnarray}
{\cal L}_{reg} &=&i\sum_{k}\bar{\psi}^k_+\gamma^\mu\left(\pam -ig\Amnl
   \snl\right)\psi^k_+ +\nonumber\\
&+&i\sum_{r}\bar{\psi}^r\gamma^\mu\left(\pam -ig\Amnl\snl\right)\psi^r +
\label{F10}\\
&+&i\sum_{s}\bar{\phi}^s\gamma^\mu
\Gamma_{11}\left(\pam -ig\Amnl\snl\right)
\phi^s -\nonumber\\
&-&\left(\sum_{r}\frac{M_r}{2}\bar{\psi}_rC_DC\Gamma_{11}
   \bar{\psi}^T_r -\sum_{s}\frac{M_s}{2}\bar{\phi}_sC_DC\bar{\phi}^T_s
   +\,\mbox{h.c.}\right)\,.\nonumber
\end{eqnarray}
Here $\psi_r$ are the fermionic P-V fields and $\phi_r$
are the bosonic ones.

The Pauli-Villars conditions
\begin{eqnarray}
2\sum_{r}C_r -2\sum_{s} C_s +2 &=&0\label{F11}\\
\sum_r C_rM^2_r -\sum_s C_s M^2_s &=&0\label{F12}
\end{eqnarray}
are assumed.
In these equations $C_t$ is the number of the P-V fields
with the mass $M_t$. The last term in the eq.(\ref{F11})
is 2 because in our case there are 2 generations of
the original fields giving the identical contributions to
the spinorial loops. The factor 2 multiplying $C_r$
and $C_s$ is due to the presence of the P-V fields of both
chiralities giving the identical contribution to the
divergent diagrams.

The propagators generated by the Lagrangian (\ref{F10})
look as follows
\begin{eqnarray}
S_{\bps^+\ps^+}&=&S_{\bps^-\ps^-}=S_{\bph^+\ph^+}=S_{\bph^-\ph^-}=
\frac{\hat k}{k^2+M^2_r}\,,\nonumber\\
S_{\bps^-\bar{\psi}^+}&=&
                        S_{\ps^+\ps^-}=S_{\bph^-\bph^+}=S_{\ph^+\ph^-}=
\frac{M_rC_DC\Gamma_{11}}{k^2+M^2_r}\,.\label{F13}
\end{eqnarray}

One sees that eqs.(\ref{F10}-\ref{F13})
define a standard Pauli-Villars
regularization. If the conditions
(\ref{F11}-\ref{F12}) are fulfilled
all spinor loops are finite. At the same time the regularized Lagrangian
(\ref{F10}) is manifestly gauge invariant. It is worthwhile to
emphasize that the number of generations being even was crucial
for the above discussion. Due to the presence of P-V
fields of both chiralities their contribution to the
divergent diagrams is twice as big as the contribution of
one generation of the original fields $\psi_+$. In the case of
the odd number of generations the last term in the eq.~(\ref{F11})
would be replaced by $(2n+1)$ and this equation cannot be
satisfied by a finite number of P-V fields. Of course, if one
allows for fractional values of $C_r$ the eq.(~\ref{F11})
may be satisfied for the odd number of generations as well.
But in this case one cannot interpret $C_r$ as the number of the
P-V fields with the mass $M_r$ and the locality of the
regularized Lagrangian may be lost.

It was shown in our paper \cite{fro} that in this case
the problem may be solved by introducing an infinite number
of P-V fields.

Obviously the regularization (\ref{F10}) works also for the
standard model. The mass terms being invariant under
SO(10) transformations are also invariant with respect to
any subgroups of SO(10) in particular $SU(3)\times SU(2)\times U(1)$.
So to get the invariant regularization of the standard model it
is sufficient to keep in the eq.~(\ref{F10}) only the gauge fields
$A^{kl}$ corresponding to the gluons and electroweak bosons
and put all other gauge fields equal to zero.

Let us generalize the regularization described above to the
case of lattice gauge models. It is well known that in the
lattice models all fermionic stats are accompanied by doubler states
due to the fact that the lattice fermion propagator
\begin{equation}
S(p) = (\sum_\mu a^{-1}\sin p_\mu a)^{-1} \label{F14}
\end{equation}
has poles not only at $p=0$, but also in the vicinity of
the points $p=(\pi a^{-1},0,0,0)$, $(\pi a^{-1},\pi a^{-1},0,0)$,
etc. Following Wilson \cite{wil} one can kill these unwanted
states by adding to the lattice action the term
\begin{equation}
\frac{\kappa}{2a}
           \sum_{x,\mu}\left(\bar\psi(x)\psi(x+a_\mu)+\bar\psi(x+a_\mu)
\psi(x)-2\bar\psi(x)\psi(x)\right)
\label{F15}
\end{equation}
This term provides all the states but one with the masses of the
order $a^{-1}$ and therefore cures the desease. In the case
of vector gauge theories it may be easily done gauge invariant
by replacing lattice derivatives by the covariant ones.
However it cannot be done for chiral gauge theories and the
Wilson term inevitably breaks the gauge invariance.

Nevertheless it will be shown that using the
regularized action of the type (\ref{F10})
one can introduce Wilson like mass terms in such a way
that all doubler states will acquire the masses of the
order of the cut-off and the gauge invariance, although
broken for a finite lattice spacing is restored in
the continuum limit. In this process no gauge non invariant
counterterms or fine tuning of the parameters is needed. The
idea is to compensate gauge noninvariant contributions of the
original fields $\psi_+$ by the corresponding contributions
of the PV fields. To do that one introduces the same Wilson
like mass term for all the fields $\psi_+,\psi_r,\phi_r$
and chooses the P-V masses $M_r\ll a^{-1}$. Then in the
vicinity of $p=0$ one can neglect the Wilson terms
recovering the gauge invariant continuum result. On the
other hand in the vicinity of $p=(\pi/a, 0,0,0)$ etc. the
leading terms in the integrands of Feynman diagrams are
zero due to the P-V conditions and the remaining
contributions vanish when $a\to 0$. The formal proof
will be given below.

The lattice action for the SO(10) model looks as follows

\begin{eqnarray}
I &=&\sum_{x,\mu,k}\Bigl[-\frac{1}{2ia}\bar\psi^+_k(x)\gamma_\mu
     U_\mu(x)\psi^+_k(x+a_\mu)-\nonumber\\
  &-&\frac{\kappa}{a}\left(\bar\psi^+_k(x)C_D\bar\psi^{+T}_k(x+a_\mu)
     -\bar\psi^+_k(x)C_D\bar\psi^{+T}_k(x)\right)+
     \mbox{h.c.}\Bigr] +\nonumber\\
  &+&\sum_{x,\mu,r}\Bigl[ -\frac{1}{2ia}\bar\psi_r(x)\gamma^\mu U_\mu(x)
      \psi_r(x+a_\mu)-\label{F16}\\
  &-&\frac{\kappa}{a}\left(\bps(x)C_D\bps^{T}(x+a_\mu)
     -\bps(x)C_D\bps^{T}(x)\right)-\frac{M_r}{2}
     \bps(x)C_DC\Gamma_{11}\bps^T(x)+\mbox{h.c.}\Bigr]
  \nonumber\\
  &+&\sum_{x,\mu,r}\Bigl[ -\frac{1}{2ia}\bar\phi_r(x)\gamma^\mu
      \Gamma_{11}U_\mu(x)
      \phi_r(x+a_\mu)-\frac{1}{2ia}{\btph}(x)\gamma_\mu
  \Gamma_{11}U_\mu(x)\tilde\phi_r(x+a_\mu)-\nonumber\\
  &-&\left(\btph (x)C_D\bph^T(x+a\mu)+
      \btph (x+a_\mu)C_D\bph^T(x)-2
      \btph (x)C_D\bph^T(x)\right)+\nonumber\\
  &+&\frac{M_r}{2}\left(\bph(x)C_DC\bph^T(x)+
     \btph(x)C_DC\btph^T\right)+
     \mbox{h.c.}\Bigr]\nonumber
\end{eqnarray}

In this equation the index $k$ as before numerates generations.
For the moment we shall take $k=1,2$.
\begin{equation}
U_\mu =\exp\bigl\{iga\sigma^{ij}A^{ij}_\mu\bigr\}
\label{F17}
\end{equation}
All other notations are the same as in eq.~(\ref{F10}) except for
the new set of bosonic P-V fields $\tilde\phi_r$. The
fields $\tilde\phi_r$ are necessary to make a nonzero mass
term $\bar\phi C_D\overline{\tilde{\phi^T}}$.

In our model there are several different dimensional parameters
like $M_r,a^{-1}$, which in the continuum limit become infinite.
It is convenient to introduce one fixed mass scale $\lambda$
and take all other masses to be proportional to
$\lambda\colon a^{-1}=\lambda N$, $M_r=\lambda N^\delta$,
etc. The continuum
limit corresponds to $N\to\infty$. In the following we assume
that $M_r\ll a^{-1}$, i.e.\ $\delta <1$. More precise condition
will be specified below.

One sees that the action (\ref{F16}) is nothing but a
discretization of the gauge invariant continuum Lagrangian
(\ref{F10}) except for the presence of the Wilson mass terms
breaking the gauge invariance. Due to these terms each generation of
the original fields $\psi_k$ has only one massless state.
Correspondingly each $P-V$ field $\psi_r,\phi_r$ describes one
state with the mass $M_r$ and 15 doublers with the masses
$\sim\kappa a^{-1}$. We shall see that when $a\to 0$
the contribution of the doubler states vanishes
and hence we are left with the same set of the
Feynman rules as the one defined by the manifestly gauge
invariant Lagrangian (\ref{F10}).

The action (\ref{F16}) generates the propagators of the same
type as given by the eq.~(\ref{F13}) and the additional
propagators $\bar\psi_\pm\bar\psi_\pm$, $\bar\phi_\pm\bar\phi_\pm$,
$\tilde\phi_\pm\tilde\phi_\pm$.

They look as follows
\begin{eqnarray}
S_{\bar\psi^k_+\psi^k_+} &=&\frac{\hat s}{s^2+m^2}\,,\label{F18}\\
S_{\bps^+\ps^+}&=&S_{\bps^-\ps^-}=S_{\bps^+\ps^+}=
S_{\btph^+\tph^+} =-S_{\bph^-\ph^-}=
-S_{\btph^-\tph^-}= \frac{\hat{s}}{s^2+m^2+M^2_r}\label{F19}\\
S_{\bar\psi_k^+\bar\psi_k^+}&=&S_{\psi_k^+\psi_k^+}=
\frac{C_D m}{s^2+m^2+M^2_r}\,,\label{F20}\\
S_{\bps^+\bps^+}&=&S_{\ps^+\ps^+}=S_{\bps^-\bps^-}=S_{\ps^-\ps^-}=
S_{\bph^+\tph^+}=S_{\bph^-\tph^-}=
\frac{C_D m}{s^2+(m^2+M^2_r)}\,,\label{F21}\\
S_{\bps^-\bps^+}&=&S_{\ps^+\ps^-}=S_{\bph^-\bph^+}=
S_{\btph^-\btph^+}=S_{\ph^+\ph^-}=S_{\tph^+\tph^-}=
\frac{M_rC_D C\Gamma_{11}}{s^2+(m^2+M^2_r)}\,,\label{F22}
\end{eqnarray}
Here
\begin{eqnarray}
s_\mu &=&a^{-1}\sin (p_\mu a)\,,\label{F23}\\
m&=&\kappa a^{-1}\sum_\mu (1-\cos (p_\mu a))\,.\label{F24}
\end{eqnarray}

Let us show that the doublers contribution vanishes in
the limit $a\to 0$. Consider for example the polarization
operator $\Pi_{\mu\nu}$. It includes the contributions of
all the fields $\psi_k,\ps,\ph,\tph$
and consists of the different pieces corresponding to the
different types of the propagators (\ref{F18}-\ref{F22})
entering the diagram.

The generic form of $\pimn$ is:
\begin{equation}
\Pi_{\mu\nu}^{(ij)(kl)}(k) =
                            \int\limits^{\frac{\pi}{a}}_{-\frac{\pi}{a}}
Tr\lbrack V_\mu^{ij}(p,q)S(p)V_\nu^{kl}(p,q)S(q)\rbrack d^4 p\quad ,
\quad  p + q + k = 0 \label{F25} \\
\end{equation}
where $S(p)$ stands for one of the propagators (\ref{F18}-\ref{F22})
and $V_\mu$ is the interaction vertex
\begin{equation}
V_\mu^{ij} =g\gm \frac{(1+\g)}{2}\sij \frac{(1+\Gamma_{11})}{2}
\cos\left[\frac{1}{2}(p-q)_\mu a\right]\,.\label{F26}
\end{equation}
We separate the integration domain in eq.~(\ref{F25}) into two
parts $V_{\mbox{in}}, V_{\mbox{out}}$, defined as follows
\begin{equation}
V_{\mbox{in}}\colon\vert p\vert <\lambda N^\gamma\ll a^{-1};
V_{\mbox{out}}\colon\vert p\vert >\lambda N^\gamma\,,\,
\gamma <\frac{1}{2}\,.
\label{F27}
\end{equation}

In the domain $V_{\mbox{in}}|pa|\ll 1$ and one can use the
expansion over $(pa)$. We shall show that the integral over
$V_{\mbox{in}}$ in the limit $a\to 0$ coincides with the
corresponding integral generated by the manifestly gauge
invariant continuum Lagrangian (\ref{F10}). Consider firstly
the diagrams including the propagators (\ref{F18}-\ref{F19}).
\begin{equation}
\pimn^{(a)}\sim g^2\int\frac{
Tr\Bigl[\gm\sigma^{ij}\hat p\gn\sigma^{kl}
(\hat p+\hat k)\Bigl(\frac{1+\g}{2}\Bigr)\Bigl(\frac{1\pm\Gamma_{11}}{2}
\Bigr)\Bigr]}
{ \lbrack p^2+\kappa^2a^2p^4+M^2_r\rbrack
\lbrack(p+k)^2+\kappa^2(p+k)^4
a^2+M^2_r\rbrack}d^4 p\,.
\label{F28}
\end{equation}
Expanding the denominator in terms of $(pa)$ one gets
\begin{eqnarray}
\pimn^{(a)}\simeq g^2\int Tr\left[\gm\sij\hat p\sigma^{kl}\gn
(\hat p+\hat k)\Bigl(\frac{1+\g}{2}\Bigr)\Bigl(\frac{1\pm\Gamma_{11}}{2}
\Bigr)\right]\,\nonumber\\
\cdot\left[\frac{1}{\lbrack p^2+M^2_r\rbrack
\lbrack (p+k)^2+M^2_r\rbrack }-
\frac{\kappa^2p^4a^2}{\lbrack p^2+M^2_r\rbrack^2
\lbrack(p+k)^2+M^2_r\rbrack}
+\ldots\right] d^4p\,. \label{F29}
\end{eqnarray}

The first term in this expression coincides exactly with the
continuum expression generated by the Lagrangian (\ref{F10}).
The next terms are majorated by
$\lambda^2N^{4\gamma-2}\sim a^\varepsilon,
\varepsilon >0$ and vanish in the limit $a\to 0$.

The diagrams including the propagators (\ref{F22})
are analyzed in the same way:
\begin{eqnarray}
& &\pimn^{(b)}\sim g^2\int\limits_{V_{\mbox{in}}}
Tr\left[\gm\sij\gn\sigma^{kl}\Bigl(\frac{1+\g}{2}\Bigr)
\Bigl(\frac{1\pm\Gamma_{11}}{2}\Bigr)\right]\,.\nonumber\\
&\cdot &\left\{\frac{M^2_r}{\lbrack p^2+M^2_r\rbrack
\lbrack (p+k)^2+M^2_r\rbrack}-
\frac{\kappa^2p^4a^2M^2_r}{\lbrack p^2+M^2_r\rbrack^2
\lbrack (p+k)^2+M^2_r\rbrack}
+\ldots\right\}d^4p\,.\label{F30}
\end{eqnarray}
Again the first term conicides with the corresponding
continuum expression and the next terms are majorated by
$N^{2\gamma -2}M^2_r$ and vanish in the limit
$a\to 0$, if $M_r=\lambda N^\delta$, $\delta\leq\gamma$.

Finally the contribution of the propagators (\ref{F20}-\ref{F21})
is proportional to the integral
\begin{equation}
\pimn^c\sim\int\limits_{V_{\mbox{in}}}g^2d^4p
\frac{\kappa^2a^2p^4d^4p}{[p^2+\kappa^2a^2p^4+M^2_r][(p+k)^2+\kappa^2a^2
       (p+k)^4+M^2_r]}
\label{F31}
\end{equation}
and vanishes in the limit $a\to 0$.

Obviously the same arguments may be repeated for the
spinorial loops with three and more external lines. For all
these diagrams the integrals over $V_{\mbox{in}}$
coincide in the limit $a\to 0$ with the corresponding integrals
generated by the manifestly gauge invariant Lagrangian (\ref{F1}).

Now we shall show that the integrals over $V_{\mbox{out}}$
do not contribute at all to the continuum limit. Again consider
as an example the polarization operator $\pimn$.
The sum of the diagrams contributing to $\pimn^{(a)}$
looks as follows
\begin{eqnarray}
& &\pimn^{(a)}=\int\limits_{V_{\mbox{out}}}\lbrace Tr\lbrack
\gm\sigma^{ij}\cos(\ha(p-q)_\mu a)\hat s(p)\gn\sigma^{kl}
\cos(\ha (p-q)_\nu a)\label{F32}\\
& &\cdot\hat s(-(p+k))\Bigl(\frac{1+\gamma_5}{2}\Bigr)
\Bigl[2\Bigl(\frac{1+\Gamma_{11}}{2}\Bigr)(s^2(p)+m^2(p))^{-1}
(s^2(p+k)+m^2(p+k))^{-1} \nonumber\\
& &+\sum_{r,\pm}\Bigl(\frac{1\pm\Gamma_{11}}{2}\Bigr)
(s^2(p+k)+m^2(p+k)+M^2_r)^{-1}(s^2(p)+m^2(p)+M^2_r)^{-1}\nonumber\\
& &-2\sum_{r,\pm}\Bigl(\frac{1\pm\Gamma_{11}}{2}(s^2(p+k)+m^2(p+k)
+M^2_r)^{-1}(s^2(p)+m^2(p)+M^2_r)^{-1}\rbrack\rbrace d^4p\nonumber
\end{eqnarray}

Here the first term describes the contribution of the two generations
of the original fields $\psi^+_k$ (hence the factor 2).
The second term describes the contribution of the fermionic P-V
fields of both chiralities (hence the summation over $\pm$).
The last term describes the contribution of the bosonic P-V fields
(the factor 2 is due to the presence of the two sets of bosonic
fields $\phi_s,\tilde\phi_s$).

As we have already discussed the part proportional to $\Gamma_{11}$
vanishes as
\begin{equation}
Tr(\sigma^{ij}\sigma^{kl}\Gamma_{11})=0\,.
\label{F33}
\end{equation}
Therefore the summation over $\pm$ simply doubles the
individual contribution.

In the domain $\mid p\mid >\lambda N^\gamma$ the integrand
in eq.~(\ref{F32}) may be expanded in terms of $M^2_r$.
The zero order term is proportional to
\begin{equation}
(2+2\sum_r C_r-4\sum_s C_s)\,,
\label{F34}
\end{equation}
where $C_r$ is the number of fermionic P-V fields with
the mass $M_r$ and $C_s$ is the number of the bosonic
P-V fields with the mass $M_s$. Using the Pauli-Villars
conditions one can make this sum equal to zero.

The first order term is proportional to
\begin{equation}
\sum_r C_r M^2_r-2\sum_s C_sM^2_s\,,
\label{F35}
\end{equation}
which again may be done equal to zero by P-V conditions.
The remaining terms are majorated by $a^2M^4_r$ and vanish
in the limit $a\to 0$.

The same reasoning may be applied to the diagrams
$\pimn^{(b)}$ and $\pimn^{(c)}$. For example the
integrand in the diagram $\pimn^{(c)}$ is proportional to
\begin{eqnarray}
& &2m^2(p)\biggl\{\frac{1}{\lbrack s^2(p)+m^2(p)\rbrack
\lbrack s^2(p+k)+m^2(p+k)\rbrack}
+\nonumber\\
&+&\sum_r\frac{1}{\lbrack s^2(p)+m^2(p)+M^2_r\rbrack
\lbrack s^2(p+k)+m^2(p+k)+M^2_r\rbrack}
-\label{F36}\\
&-&\sum_r\frac{1}{\lbrack s^2(p)+m^2(p)+M^2_r\rbrack
\lbrack s^2(p+k)+m^2(p+k)+M^2_r\rbrack}\biggr\}
\nonumber
\end{eqnarray}
Expanding it in terms of $M^2_r$ one sees that the first two
terms are zero due to the P-V conditions and the remaining
terms are majorated by $a^2M^4_r$.

In the lattice case there are also additional tadpole
diagrams which are absent in the continuum theory. They arise
when one expand $U_\mu (x)$ in terms of $A_\mu$
and considers the higher order terms. These diagrams are analysed
exactly in the same way and one sees that they do not contribute to
the continuum limit.

Generalization to the spinor loops with more than two
external lines is absolutely straightforward. Integrals
over $V_{\mbox{out}}$ vanish in the limit $a\to 0$ for all
spinor loops.

Therefore we proved that for small lattice spacing
the Feynman rules for the spinor loops which follow from
the lattice action (\ref{F16}) are identical to the manifestly
gauge invariant rules generated by the continuum Lagrangian
(\ref{F10}). To remove ultraviolet divergencies one needs only
the usual gauge invariant counterterms.
\bigskip
\section{Regularization of the chiral models with an odd number
of generations.}

Now we pass to the discussion of the SO(10) model
with an odd number of generations. Again we start
by reminding the procedure for the continuum model
\cite{fro}.

The difficulty with the regularization of the odd number generation
model is due to the fact that the P-V fields always enter with
both chiralities, and one cannot satisfy the P-V condition
\begin{equation}
2\sum C_r-2\sum C_s +(2n+1)=0
\label{F37}
\end{equation}
by a finite number of P-V fields.

Instead it was proposed in the paper \cite{fro} to introduce
an infinite system of P-V fields $\psi_r$ with the masses
$M_r=M\mid r\mid$ and Grassmanian parity $\varepsilon (\psi_r)=
(-1)^{r-1}$. The index $r$ changes from $-\infty$ to $+\infty$,
$r=0$ corresponds to the original field $\psi_+$, positive $r$
numerating the positive chirality P-V spinors, negative $r$
numerating the negative chirality ones.

As it was discussed above the contribution of the positive
and negative chirality spinors to the divergent diagrams are
equal. For definites we shall take the number of
generations equal to 1.

The propagators are given by the eqs.(\ref{F13}). The integral
corresponding to the diagrams $\pimn^{(a)}$ looks as follows
\begin{equation}
\pimn^{(a)}(k)=\int dp\frac{\sum{(-1)^r}Tr\lbrack\gm\sij\hat p\gn
\sigma_{kl}(\hat p+\hat k)\Bigl(\frac{1+\gamma_5}{2}\Bigr)\rbrack}
    {\lbrack p^2-M^2r^2\rbrack\lbrack (p+k)^2-M^2r^2\rbrack}\,.
\label{F38}
\end{equation}
Here we omitted the term proportional to $\Gamma_{11}$ because its
contribution is equal to zero as we have shown above.

The leading term in the integrand for $p\to\infty$ is
\begin{equation}
\sim Tr\lbrack\gamma_\mu\sigma^{ij}\hat p\gn\sigma^{kl}(\hat p+\hat k)
\Bigl(\frac{1+\gamma_5}{2}\Bigr)\rbrack\sum^{+\infty}_{r=-\infty}
\frac{(-1)^r}{(p^2+M^2r^2)^2}\,.
\label{F39}
\end{equation}
Using the representation
\begin{equation}
\sum^{+\infty}_{r=-\infty}\frac{(-1)^r}{(p^2+M^2r^2)^2}=
-\frac{\partial}{\partial p^2}\sum^{+\infty}_{r=-\infty}
\frac{(-1)^r}{p^2+M^2r^2}\,,
\label{F40}
\end{equation}
one can do the summation over $r$ explicitely
\begin{equation}
\sum^{+\infty}_{r=-\infty}\frac{(-1)^r}{p^2 +M^2r^2}=
\frac{\pi}{MR\sin h(\frac{\pi R}{M})}\,,
R^2=p^2_0+p^2_1+p^2_2+p^2_3\,.
\label{F41}
\end{equation}
One sees that the leading term in the integrand decreases
exponentially.

Next to leading terms are analyzed analogously by expanding the
integrand in the eq.(\ref{F38}) in a series over $k_\mu$.
The integrands of the corresponding terms are proportional to
\begin{equation}
\sum^{-\infty}_{r=-\infty}\frac{(-1)^r}{(p^2+M^2r^2)^n}\,,
n>2
\label{F42}
\end{equation}
and can be calculated by differentiating the eq.(\ref{F41})
with respect to $p^2$. All these terms are decreasing exponentially
and therefore the integral (\ref{F38}) is convergent.

The diagrams $\pimn^{(b)}$ including the propagators $\psi_r^+\psi_r^-$
are treated in the same way. For example the integrand for $\pimn^{(b)}$
is proportional to
\begin{equation}
\sum^\infty_{r=-\infty} (-1)^r
\frac{M^2r^2}{\lbrack (p+k)^2+M^2r\rbrack (p^2+M^2r^2)}\,.
\label{F43}
\end{equation}
Representing $M^2_r$ in the numerator as $(M^2r^2+p^2)-p^2$ one
reduces the problem of summation over $r$ to the case
considered above. The corresponding functions decrease exponentially
providing the convergence of the integrals.

Generalization to the diagrams with more than two external lines
is obvious: all these diagrams correspond to the convergent integrals.

To analyze the convergence of the regularized diagrams it is not
necessary in fact to expand them in series over $k_\mu$.
Instead one can use the Feynman representation
\begin{equation}
\frac{1}{[p^2+M^2_r](p+k)^2+M^2_r]}=\int\limits^1_0
\frac{d\alpha}{[(1-\alpha)(p^2+M^2_r)+\alpha[(p+k)^2+M^2_r]]^2}\,.
\label{F44}
\end{equation}
Shifting the integration variables $p_\mu$,
$p_\mu\to p_\mu -\alpha k_\mu$, one can write this expression
in the standard form
\begin{equation}
\frac{1}{\lbrack p^2+\alpha (1-\alpha)k^2+M^2_r\rbrack^2}\,.
\label{F45}
\end{equation}
Now the summation over $r$ can be done explicitly for arbitrary
$k$ leading to the same conclusion about the convergence of the
integrals. The representation (\ref{F45}) is also
useful for practical calculations.

Therefore the infinite set of P-V fields with the Majorana
masses provides a gauge invariant regularization in the case
of the odd number of generations as well.

Generalization to the lattice models goes along the same lines
as for the case of the even number of generations. The lattice
Lagrangian is given by the eq.(\ref{F16}) where now there is
no summation over $k, k=1$.

The propagators are given by the eqs.(\ref{F18}-\ref{F22}).
To get the analog of the eq.(\ref{F38}) we choose the
following set of P-V fields. The fields $\psi_{(\pm 1)}$,
$\psi_{(\pm 2)},\ldots\psi_{(\pm n)}$ are the fermions
with the masses $M\vert r\vert$, and the fields $\phi_{(\pm 1)}$,
$\phi_{(\pm 3)},\ldots \phi_{(\pm (2n+1))}$; $\tilde\phi_{(\pm 1)}$,
$\tilde\phi_{(\pm 3)},\ldots\tilde\phi_{(\pm (2n+1)}$ are the
bosons with the masses $M\vert r\vert$. As before positive $r$
correspond to the positive chirality fields and negative $r$
correspond to the negative chirality ones. With this choice
the contribution of the fields with the number $r$ to the integrand
of the polarization operator $\pimn$ is proportional to
\begin{equation}
\frac{(-1)^r}
{\lbrack s^2(p)+m^2(p)+M^2r^2\rbrack
\lbrack s^2(p+k)+m^2(p+k)+M^2r^2\rbrack}\,.
\label{F46}
\end{equation}

The analog of the eq.(\ref{F38}) for $\pimn^a$ now looks
as follows
\begin{eqnarray}
\pimn^{(a)}&& =\int\limits^{\frac{\pi}{a}}_{-\frac{\pi}{a}}d^4p
\sum^{+\infty}_{r=-\infty}(-1)^r  \cdot\nonumber \\
           &\cdot & \frac{Tr\lbrack\gm\sigma^{ij}\hat s(p)
      \cos\lbrack\ha (p-q)a_\mu\rbrack
      \gm\sigma^{kl}\hat s(p+k)\cos\lbrack\ha (p-q)a_\nu\rbrack
      \Bigl(\frac{1+\gamma_5}{2}\Bigr)\rbrack}
     {\lbrack s^2(p)+m^2(p)+M^2r^2\rbrack
     \lbrack s^2(p+k)+m^2(p+k)+M^2r^2\rbrack}
\label{F47}
\end{eqnarray}
As before we separate the integration domain
into $V_{\mbox{in}}\colon\mid p\mid <\lambda N^\gamma\ll a^{-1}$;
$\gamma <\ha$, $V_{\mbox{out}}\colon\mid p\mid >\lambda N^\gamma$.
In the domain $V_{\mbox{in}}$ we can expand the integrand in terms
of $(pa)$. In this way one gets the expression which differs
from the continuum expression (\ref{F38}) only by
the presence of the Wilson mass term $m^2\approx\kappa^2a^2p^4$:
\begin{equation}
\pimn^{(a)}(k)=\int\limits_{V_{\mbox{in}}}d^4p
\sum^\infty_{r=-\infty}(-1)^r
\frac{Tr\Bigl[\gm\sigma^{ij}\hat p\gn\sigma^{kl}
      (\hat p+\hat k)\Bigl(\frac{1+\gamma_5}{2}\Bigr)\Bigr]}
     {\lbrack p^2+\kappa^2p^4a^2+M^2r^2\rbrack
     \lbrack(p+k)^2+\kappa^2(p+k)^4a^2+M^2r^2\rbrack}
\label{F48}
\end{equation}
In the domain $\mid p\mid<\lambda N^\gamma$ the Wilson term
$\kappa^2 a^2 p^4\ll p^2+M^2 r^2$. The series in eq. (\ref{F48})
converges for any $p^2$ and one can neglect the Wilson term,
recovering the continuum expression (\ref{F38}).
In the domain $V_{\mbox{out}}$ we can expand the integrand of
the eq.~(\ref{F47}) in a series over $k^\mu$. The zero order term
is
\begin{equation}
\pimn^{(a)}(k)=\int_{V_{\mbox{out}}} d^4p\sum^{+\infty}_{r=-\infty}
(-1)^r
\frac{Tr\left[\gm\sigma^{ij}\hat s(p)\gn\hat s(p)\cos(pa_\mu)
      \cos(pa_\nu)\Bigl(\frac{1+\g}{2}\Bigr)\right]}
     {\left[s^2(p)+m^2(p)+M^2r^2\right]^2}
\label{F49}
\end{equation}
The summation over $r$ is done in the same way as in
the continuum case
\begin{equation}
\sum^{+\infty}_{r=-\infty}
\frac{(-1)^r}{[s^2(p)+m^2(p)+M^2r^2]}=-\frac{\partial}{\partial s^2}
\left(\frac{\pi}{\sqrt{M^2(s^2+m^2)}\sin h
           \left(\frac{\pi\sqrt{s^2+m^2}}{M}\right)}\right)\,.
\label{F50}
\end{equation}
For small $a$ this expression decreases exponentially. Therefore
the integral over $V_{out}$ is equal to zero in the limit
$a\to 0$. Next order terms in the series over $k^\mu$ are
analyzed in the same way.
These arguments are easily extended to any spinor loop.
The corresponding integrals can be written in the form
\begin{equation}
I_n=\int^{\frac{\pi}{a}}_{-\frac{\pi}{a}} d^4p
\sum^{+\infty}_{r=-\infty}\sum^{n-1}_{l=0}
\frac{A_l(p,Q,M_r)}{s^2(p+Q_l)+m^2(p+Q_l)+M^2r^2}
\label{F51}
\end{equation}
\begin{eqnarray*}
Q_l=k_1+\cdots + k_l\,.
\end{eqnarray*}
where $A_l$ is a polinomial in $M^2_r$. The summation over
$r$ can be done explicitely using eq.~(\ref{F50}).
One gets for $I_n$
\begin{equation}
I_n=\int^{\frac{\pi}{a}}_{-\frac{\pi}{a}}
\sum^{n-1}_{l=0}
\frac{\tilde A_l (P,Q)}
{\sqrt{M^2(s^2+m^2)}\sin h\left(\frac{\pi\sqrt{s+m^2}}{M}
\right)}d^4p
\label{F52}
\end{equation}
In the domain $V_{\mbox{in}}$ one can expand the integrand
in terms of $(pa)$. The Wilson term $\sim\kappa^2a^2p^4\ll p^2$
and can be neglected. Hence we recover the gauge invariant
continuum result.

In the domain $V_{\mbox{out}}, \sqrt{s^2+m^2}\gg M$ and the
integrand vanishes exponentially when $a\to 0$.

The final conclusion is that for small $a$ the expression
for the spinorial loops in our lattice model coincides with the
manifestly gauge invariant expression generated by the
continuum Lagrangian (\ref{F10}) both in the case of even
and odd number of generations. It is worthwhile to note that in
the lattice case it is not necessary to take infinite number of
P-V fields. It is sufficient to take a finite number
$N^1(a)$ of such fields which becomes infinite when
$a\rightarrow 0$. Indeed the series in eq.(\ref{F52}) is convergent
for any $p$ and the integration domain is finite for
$a\neq 0$. Therefore choosing $N^1$ big enough
one can always make the contribution of the remaining
fields $\sum^\infty_{|n|=N_1}$ as small as one wishes.

In the continuum limit the number of P-V fields becomes
infinite and one gets the same gauge invariant result.

Up to now we discussed only the spinor loops. There are other
diagrams to be worried about, in particular spinor particle
self energy diagrams. In the continuum case we assumed
that the higher covariant derivative regularization was
introduced for the gauge fields. It happens that in the
lattice model such a regularization is also needed. Although
for a finite lattice spacing the self energy diagrams
are finite without any additional regularization in the
continuum limit they may require noninvariant counterterms
like fermion mass renormalization. Higher covariant
derivatives for gauge fields cure this decease.

One modifies the lattice Yang-Mills action as follows:
\begin{eqnarray}
& &S_w=-
\frac{1}{g^2}\sum Tr (U^{\mu\nu}+(U^{\mu\nu})^+)\rightarrow
\label{F53}\\
&\rightarrow&-\frac{1}{g^2}\sum\Bigl\{Tr~U^{\mu\nu}
+\frac{(\Lambda^2a^2)}{g^2}\Bigl[ Tr~U^{\mu\nu}(x)U_\rho(x)
       (U^{\mu\nu}(x+a_\rho))^+U^+_\rho (x)\nonumber\\
& &\qquad\qquad\qquad  -Tr~U^{\mu\nu}(x)(U^{\mu\nu}(x))^+
\Bigr]\Bigr\}+\mbox{h.c.}\nonumber
\end{eqnarray}
where
\begin{equation}
U_{\mu\nu}=U_\mu(x)U_\nu(x+a_\mu)U^+_\mu(x+a_\nu)U^+_\nu(x)
\label{F54}\\
\hfill\mbox{(no trace).}\hfill\\
\end{equation}
It leads to the following modification of the gauge field propagator
(assuming a diagonal gauge):
\begin{eqnarray}
& &G(p)=\left[
\frac{1}{a^2}\Bigl(\sum_\mu\cos(p_\mu a)-1\Bigr)\right]^{-1}
\rightarrow\label{F55}\\
&\rightarrow&\left[a^2\Bigl(\sum_\mu(\cos(p_\mu a)-1)+\Lambda^2
\Bigl(\sum_\mu(\cos (p_\mu a)-1)\Bigr)\Bigl(\sum_\nu(\cos(p_\nu a)-1)
\Bigr)\right]^{-1}\,.\nonumber
\end{eqnarray}
Choosing $a^{-\frac{3}{2}}\ll\Lambda\ll a^{-2}$ we can suppress
the gauge field propagator so that the doublers contribution
to the fermion self energy and all other diagrams vanishes
in the limit $a\to 0$. To show it one again separates the
integration domain into $V_{\mbox{in}}$ and $V_{\mbox{out}}$
and proves that the integrals over $V_{\mbox{in}}$ coincide
with the gauge invariant continuum limit and the ingetrals
over $V_{\mbox{out}}$ are zero in this limit. Let us consider
as an example the self energy diagram for the original fermion field
including the propagator $\bar\psi^+\bar\psi^+$. In the domain
$V_{\mbox{in}}$ it is given by the integral
\begin{equation}
\sum\limits_{\mbox{in}}\sim\int\limits_{V_{\mbox{in}}}
\frac{\kappa ap^2d^4p}
     {[p^2+\kappa^2a^2p^4][(p+k)^2+\Lambda^2a^4(p+k)^4]}d^4p.
\label{F56}
\end{equation}
This diagram is majorated by $N^{2\gamma-1}$ and as $\gamma <\ha$
vanishes in the limit $a\to 0$.

The integral over $V_{\mbox{out}}$ looks as follows
\begin{equation}
\sum_{\mbox{out}}\sim\int_{V_{\mbox{out}}}
\frac{m(p)\Bigl[\cos\ha (p-q)_\mu a\Bigr]^2}
{[s^2(p)+m^2(p)]\Bigl[a^{-2}\sum_\mu(\cos(p_\mu a-1)+\Lambda^2
\Bigl(\sum_\mu(\cos (p_\mu a)-1)\Bigr)^2\Bigr]}
\label{F57}
\end{equation}
In the limit $a\to 0$, $\sum_{\mbox{out}}\leq\Lambda^{-2}a^{-3}\to 0$.

Note that if the higher covariant derivatives for gauge fields
were absent ($\Lambda=0$) this term would produce an infinite
mass renormalization for the fields $\psi$.

All other diagrams including internal gauge lines may be
treated in the same way demonstrating a manifestly gauge
invariant continuum limit.

It completes the proof of the gauge invariance of our construction.
The lattice action (\ref{F16}) together with the higher
derivative regularized gauge field action (\ref{F51}) leads to
the continuum theory which is free of doublers and gauge
invariant.
\bigskip
\section{Discussion}
We demonstrated above that the problem of removing
the fermion doublers in anomaly free chiral gauge models
may be solved by introducing the lattice action defined
by the eqs.~(\ref{F16}),(\ref{F51}). This action breaks
the gauge invariance for finite lattice spacing but the
invariance is restored in the continuum limit. The only
counterterms which are needed to take a continuum limit
are the usual gauge invariant counterterms. From the point of view
of calculations the procedure described above is quite
simple in the case of even number of generations

In this case it is essentially a discretization of the usual
Pauli-Villars method. So in many practical calculations one probably
can neglect the third generation which contains heavy
particles and use this simple procedure.

It goes without saying that the arguments presented in this
paper used the weak coupling expansion. It would be of
great interest to check their validity in nonperturbative
calculations.

\subsection*{Acknowledgements}
A.A. Slavnov would like to express his gratitude to D. Maison and
W. Zimmermann for hospitality extended to him at Max-Planck Institute for
physics in Munich, G. Mack for hospitality at the University of Hamburg,
and Volkswagen-Stiftung for financial support. He is also grateful to
M. L\"uscher and I. Montvay for helpful discussions.
S.A.Frolov thanks P.K. Mitter for hospitality at Laboratoire de Physique
Theorique et des Hautes Energies de Universite Pierre et Marie Curie
in Paris.

\end{document}